# A new search for anomalous neutrino oscillations at the CERN-PS

B. Baibussinov[a], E. Calligarich[b] S. Centro[a], D. Gibin[a], A. Guglielmi[a],
F. Pietropaolo[a], C. Rubbia[c,*] and P. Sala[d]

[a] *Dipartimento di Fisica dell'Università e Sezione INFN di Padova, via Marzolo 8, I-35131 Padova, Italy*
[b] *Sezione INFN di Pavia, via Bassi 6, I-27100 Pavia, Italy*
[c] *CERN, European Laboratory for Particle Physics, CH-1211 Geneve 23, Switzerland*
[d] *Sezione INFN di Milano, via Celoria 2, I-20123 Milano, Italy*

*E-mail (\*corresponding author)*: carlo.rubbia@cern.ch

ABSTRACT: The LSND experiment at LANSCE has observed a strong 3.8 σ excess of $\bar{\nu}_e$ events from an $\bar{\nu}_\mu$ beam coming from pions at rest. If interpreted as due to neutrino oscillations, it would correspond to a mass difference much larger and inconsistent with the mass-squared differences required by the standard atmospheric and long-baseline neutrino experiments. Therefore, if confirmed, the LSND anomaly would imply new physics beyond the standard model, presumably in the form of some additional sterile neutrinos.

More recently the MiniBooNE experiment at FNAL-Booster has further searched for the LSND anomaly both for $\nu_e$ and $\bar{\nu}_e$ with a horn focused neutrino beam from protons of the FNAL Booster. The anomaly would then manifest itself as a difference between the experimental observations and predictions based on Monte Carlo simulations supported by the rather limited existing neutrino cross sections. Above 475 MeV, the $\nu_e$ result is excluding the LSND anomaly to about 1.6 σ but it introduces an unexplained, new 3.0 σ anomaly at lower energies, down to 200 MeV. The $\nu_e$ data have so far an insufficient statistics to be conclusive with LSND's $\bar{\nu}_e$.

The present proposal at the CERN–PS is based on two strictly identical LAr-TPC detectors in the "near" and "far" positions, respectively at 127 and 850 m from the neutrino (or antineutrino) target and focussing horn, observing the electron-neutrino signal. In this way, all cross sections and experimental biases cancel out. In absence of oscillations the two event distributions must be identical, the electron neutrino radial and energy distributions being extremely similar in the two positions. On the other hand, any difference of the event distributions due to the position between the two detectors should be attributed to the possible existence of the LSND-like anomalies. If the presence of a distance dependent oscillation is detected by the experiment, it would be possible to observe the actual oscillation as a function of ($\Delta m^2$ – $\sin^2 2\theta$) plane and therefore to provide a measure of the mixing angle and of the mass difference due to the LSND anomaly.

This project will benefit from the already developed technology of ICARUS T600, well tested on surface in Pavia, without the need of any major R&D activity and without the added problems of an underground experiment (CNGS2). The superior quality of the Liquid Argon imaging TPC and its unique $e - \pi^0$ discrimination allow full rejection of the NC background, without efficiency loss for electron neutrino detection. In two years of exposure, the "far" detector mass of 600 tons and a reasonable utilization of the CERN-PS with the refurbished previous TT7 beam line will allow to collect about $10^6$ charged current events, largely adequate to settle definitely the LSND anomaly.

# Contents

## 1. Introduction

Neutrinos have been the origin of an impressive number of "surprises". The LEP experiments have demonstrated that the sum of the strengths of the coupling of different neutrinos is very close to 3, confirming a previous result by UA1. *But it is only assuming that neutrinos, in similarity to charged leptons, have unitary strengths that the resulting number of neutrinos is 3.*

Recent results have shown that a precise identity between neutrino and charged lepton families is by no means automatically granted. The recent observation of the extraordinarily large neutrino mixings when compared to naïve Cabibbo-like expectations from quarks, the absence of right-handed partners, the possibility of Majorana-like couplings, the extraordinary small mass spectrum and so on, leave a large number of potential features untested. The experimentally measured weak coupling strengths are only rather poorly known, leaving room for many other alternatives.

It is only because the masses of known neutrino species are so small, that their contribution to the Dark Matter of the Universe can be excluded. The situation could be altered by the additional presence of sterile neutrinos, if sufficiently massive. Therefore, the presence of massive sterile neutrinos will contribute to clarify the Dark Matter problem.

Over the last several years, a systematic search for oscillations in the ($\Delta m^2 - \sin^2 2\theta$) plane has been performed by a large number of experiments at CERN, FNAL, KEK, ISIS, LANSCE and elsewhere, in order to identify the possible presence of oscillation signals. Additional Physics beyond the Standard Model in the neutrino sector will be necessary if more than the two oscillation signals due to Atmospheric and Solar transitions were to be eventually observed.

Positive evidence for positron excess of in $\bar{\nu}_\mu$ interactions coming from muon decays at rest has been reported by the LSND experiment at LANSCE in Los Alamos [1]. They have observed (Figure 1) $\bar{\nu}_e + p \rightarrow e^+ + n$ interactions in excess of the expected contributions due to $\bar{\nu}_e$ production. The anomalous $\bar{\nu}_e$ signal (87.9 ± 22.4 ± 6.0) represents a 3.8 σ effect and it occurs for L/E distances of about 0.5 – 1.0 m/MeV. To explain this signal with neutrino oscillations requires a mass-squared difference $\Delta m^2 \approx 1$ eV$^2$. If interpreted as due to neutrino oscillations, taking into account the distance between the source and the detector, it corresponds to a mass difference much larger and inconsistent with the mass-squared differences required by the standard atmospheric and long-baseline neutrino experiments. Additional observation of neutrinos from pion decays in flight [2] has given a signal of (18 ± 6.6) events with a background of 21 events, consistent to the signal at rest but with a much weaker statistical impact. This result has remained unchallenged by all other experiments.

Many theoretical hypotheses (see for instance [3]) have been put forward to explain the signal excess observed by LSND and to accommodate it into a coherent and more general neutrino physics framework. All of them imply the presence of new physics beyond the Standard Model, in most instances with additional sterile neutrino families, whose existence is probed through their coupling with ordinary matter via neutrino oscillation, decay or CPT violation. Other more exotic scenarios have also been suggested.

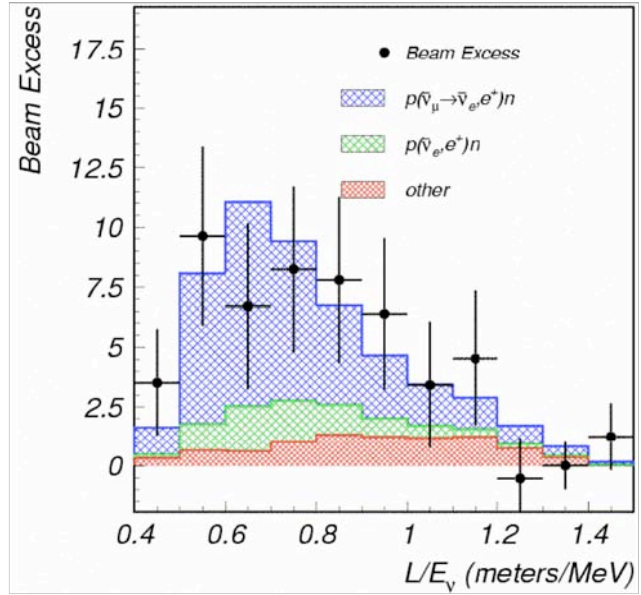

**Figure 1.** Excess of $\bar{\nu}_e$ events in a $\bar{\nu}_\mu$ beam as observed by the LSND experiment (3.8σ above the expected background) [1].

More recently the MiniBooNE experiment at FNAL-Booster has further searched for the LSND anomaly with a horn focused neutrino beam from protons of the FNAL Booster. Data have been collected with an initial beam of $\nu_\mu$ neutrinos [4] and more recently [5] with $\bar{\nu}_\mu$. In view of the unknown nature of the process, the oscillation patterns for the $\nu_\mu$ and $\bar{\nu}_\mu$ might not be the same. The 450 t liquid scintillator detector was exposed at a ~ 0.7 GeV $\nu_\mu$ beam at 550 m from the target for an integrated intensity of 6.6 $10^{20}$ pot. The LSND expectation is an additional signal over a background of 375 CC-QE counts, due to a variety of backgrounds mostly calculated with Monte Carlo predictions, like for instance $\pi^0$ production, and the intrinsic $\nu_e$ beam contamination. For energies E > 475 MeV, the result (Figure 2) is in agreement with the absence of a LSND-like oscillation (but only at 90 % level, 1.6 σ). However in the 200 – 475 MeV energy range, the assumption of no oscillations shows an excess of (128.8 ± 20.4 ± 38.3) events above a background of 415 events (3.0 σ effect) between experimental data and Monte Carlo predictions [6]. The detector has also been exposed to a $\bar{\nu}_\mu$ beam, based on 3.4 $10^{20}$ pot. Due to insufficient statistical evidence, the result (Figure 3) is however so far compatible with the LSND result. MiniBooNE relies heavily on Monte Carlo simulations based on the extremely scarce neutrino events from 40 years old bubble chambers.

In order to conclusively describe the LSND anomaly, three new experiments are under preparation: (1) the CNGS2-ICARUS experiment in the LNGS [7]; (2) MicroBooNE at FNAL [8]; (3) OscSNS at ORNL, an experiment similar to LSND but with the higher intensity spallation source (1.4 MW) planned at SNS [9].

The CNGS2-ICARUS is observing high energy neutrino oscillations after 732 km from CERN [10] in a 600 ton liquid argon time projection chamber (LAr-TPC). The oscillation pattern is much longer, characteristic of L/E ≈ 40 m/MeV and we expect to have complete mixing for a LSND anomaly. Possible backgrounds are the intrinsic $\nu_e$ contamination and electronic decays from $\nu_\mu \to \nu_\tau$ oscillations occurring with a much smaller mass difference. The expected sensitivity to LSND in the ($\Delta m^2$ – $\sin^2 2\theta$) plane is substantial but limited to larger values of $\sin^2 2\theta$ (Figure 11-top). In addition, the programme at CNGS foresees only the use of neutrino beam, while the LSND anomaly is related to antineutrino. Therefore, although this experiment provides valid information, it cannot per-se entirely exclude the whole range of possibilities for the LSND anomaly.

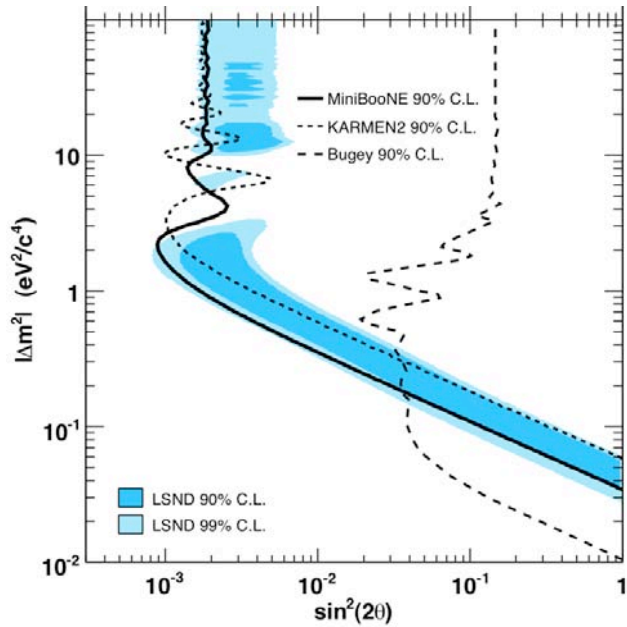

**Figure 2.** LSND allowed oscillation parameters region, compared with the negative results from KARMEN2, Bugey and MiniBooNE. The MiniBooNE exclusion plot refers to quasi-elastic neutrino events (0.475 – 3.0 GeV energy range) and is based on 6.46 x $10^{20}$ pot.

The MicroBooNE at FNAL is a LArTPC detector from the Booster and it is based, like the ICARUS experiment, on the LAr TPC technology. The fiducial mass is only about 70 ton, almost one order of magnitude smaller than the MiniBooNE and ICARUS experiments, although with respect to MiniBooNE the recognition of most of the channels is now becoming possible. This is an advantage in comparison to MiniBooNE, that was capable to positively identify only a fraction of the quasi-elastic events. However the very heavy reliance on Monte Carlo simulations still persists. The beginning of the operation of the MicroBooNE experiment is foreseen by 2012/13.

Finally the new experiment OscSNS has been proposed at ORNL with pions at rest, but with a higher intensity spallation source (1.4 MW). The detection principle is identical to the one of LSND. The new detector has essentially the same mass (800 t) as the one already used at LAMPF, but with higher PMT coverage and at a distance of ~ 60 m from the SNS beam stop at ORNL. If approved within 2009, the experiment could begin to operate around 2013 with an appearance signal and about three times the statistics of LSND for the nominal SNS intensity.

The present proposal at the CERN-PS introduces some important new features, which should allow a definitive clarification of the LSND anomaly. In the MiniBooNE and in the planned MicroBooNE experiments, any difference between the observed "electron" signal at a given distance and the Monte Carlo fitted prediction is interpreted in terms of evidence for neutrino oscillations. In reality there are large differences in MiniBooNE between data and predictions, but they do not seem to correspond to the features of a neutrino oscillation over the full observable energy interval, although some theoretical model may accommodate it [11].

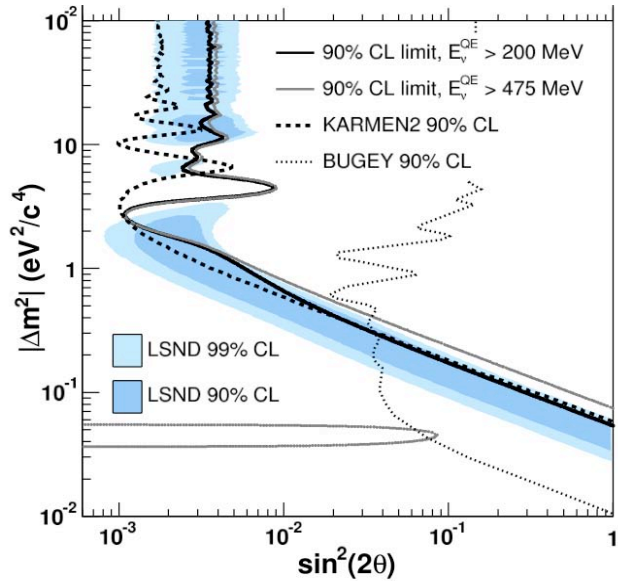

**Figure 3.** Same as in Figure 2 but the MiniBooNE exclusion plot refers to quasi-elastic anti-neutrino events (for two different energy threshold: 0.475 GeV and 0.2 GeV) and is based on $3.39 \times 10^{20}$ pot.

The present proposal is based on a new feature, the search for spectral differences of electron like specific signatures *in two identical detectors but at two different neutrino decay distances*, at the "far" and the "near" locations, respectively at 850 m and 127 m away from the source (Figure 4). Evidently, in absence of oscillations, after some beam related small spatial corrections, the two energy spectra should eventually be a precise copy of each other, independently of the specific experimental event signatures and without any need of Monte Carlo comparisons. Any resulting $\nu_e$ difference between the two locations, if observed, must be inevitably attributed to the time evolution of the neutrino species due to the presence of oscillations.

The detector is based on the LAr-TPC technology, first proposed by Carlo Rubbia in 1977, and developed by the ICARUS group since about two decades [12]. It will have an active mass comparable to the one of the ICARUS T600 detector [13], which was initially realized and tested on surface in Pavia and more recently installed in the underground area at the LNGS. The fiducial mass and the corresponding number of events will be about an order of magnitude larger than the one of the MicroBooNE LAr detector. The "bubble chamber like" observation of the neutrino events permits to identify precisely the features of individual events, while all previously reported observations have been based on the necessarily more primitive observation of Cerenkov rings produced by PM's at the surface of the detector volume and which are mostly limited to quasi-elastic events.

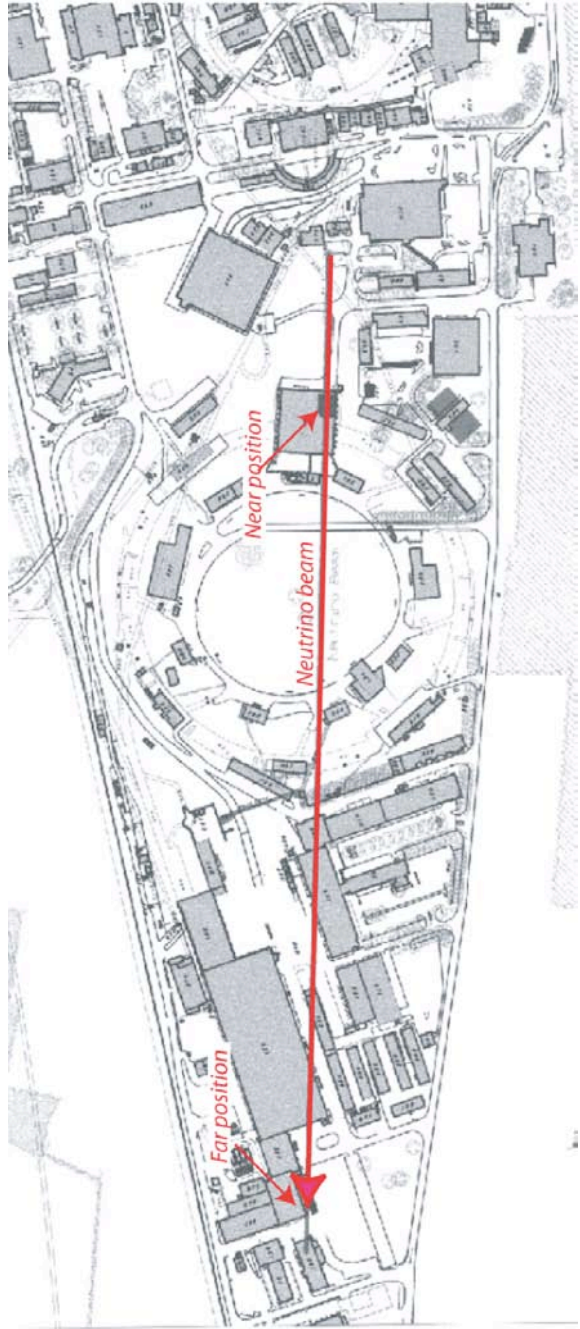

**Figure 4.** Neutrino beam from the CERN-PS. Two locations, respectively at 850 and 127 m from the target are simultaneously recorded in order to evidence possible oscillation effects.

Associated with the refurbishing of the CERN-PS low energy neutrino beam originally used by the BEBC-PS180 collaboration [14], at a distance of about 850 m from the target and for $1.25 \times 10^{20}$ pot/year, the detector of the far location will accumulate more than 600'000 unbiased, "bubble chamber like" neutrino events each year. The LAr-TPC may accumulate such a very large statistics with the very high spatial and energy resolutions of the LAr and the clear identification of individual NC and CC channels. These results will be complementary to the ones from ICARUS T600, exposed at CNGS, which will detect $\nu_\mu \to \nu_e$ oscillations at much larger L/E but with smaller statistics and therefore larger $\sin^2 2\theta_{e\mu}$, with an enhanced sensitivity toward low $\Delta m^2$ values.

Because of the evident advantages of the LAr-TPC, the exposure could also be used for precise measurements of neutrino cross-sections in the low energy range, useful to improve the future neutrino oscillation experiments in the low energy region up to a few GeV. Moreover a LAr-TPC will permit to prove on a very large statistics the identification and rejection capabilities of all the NC backgrounds in LAr, which is a fundamental item for a large LAr detector in any future long baseline $\nu_\mu \to \nu_e$ oscillation search in the ~ GeV energy range.

## 2. The physics programme

The experiment will exploit the CERN-PS neutrino beam-line, originally used by the BEBC-PS180 Collaboration and successively re-considered by the I216/P311 Collaboration [15]. The neutrino beam is a low energy $\nu_\mu$ beam, produced by 19.2 GeV protons, of intensity 1.25 $10^{20}$ pot/yr.

In order to reduce the systematic errors in the search of $\nu_\mu \to \nu_e$ oscillations, it is envisaged to build two LAr-TPC cloned from the ICARUS T600 design. Since the detector is to be run on surface, the reference detector is the one operated in Pavia in 2001 and comprehensively described in Ref. [13], to which we refer for details. As a consequence, the proposed experiment would not be subject to all the additional safety features, specific to underground operation, which apply, for istance, in the case of ICARUS T600 at LNGS.

The technology of the LAr-TPC is conceived as a tool for a completely uniform imaging with high accuracy of massive volumes. The operational principle of the LAr-TPC is based on the fact that in highly purified LAr ionization tracks can be transported practically undistorted by a uniform electric field over macroscopic distances. Imaging is provided by a suitable set of electrodes (wires) placed at the end of the drift path continuously sensing and recording the signals induced by the drifting electrons. Non-destructive read-out of ionization electrons by charge induction allows detecting the signal of electrons crossing subsequent wire planes with different orientation. This provides several projective views of the same event, hence allowing space point reconstruction and precise calorimetric measurement.

The ICARUS T600 detector consists of a large cryostat split into two identical, adjacent half-modules, hosting the LAr-TPC's, for a total active mass of about 600 t. Each TPC is equipped with three parallel planes of wires, 3 mm apart. The signals from each wire are digitized every 400 ns, corresponding to a longitudinal granularity of about 0.6 mm. All technical aspects of the system, namely cryogenics, LAr purification, read-out chambers, detection of LAr scintillation light, electronics and DAQ had been tested during the commissioning in Pavia and performed as expected. Statistically significant samples of cosmic ray events (long muon tracks, spectacular high multiplicity muon bundles, electromagnetic and hadronic showers, low energy events) were recorded. The subsequent analysis of these events, carried out in 2002-03, has allowed the development and fine tuning of the off-line tools for the event reconstruction and the extraction of physical quantities. It has also demonstrated the performance of the detector in a quantitative way, as described in a number of published papers [16,17]. This test has demonstrated the maturity of the project.

In this proposal, the far detector, at the 850 m location in the Hall 191 at CERN (BEBC Hall), will have almost the same size of the T600, to minimize the design effort, while keeping an active mass (~500 tons) suitable for the proposed oscillation search. The near detector, located in the old ISR Hall 193 at a distance of 127 m from the neutrino beam target, will have an identical performance. Its event rate is about a factor 45 larger due to the closer distance from the target; even a much smaller fiducial mass will not affect the overall statistical error of the measure. An active mass of 150 ton is assumed as a reference.

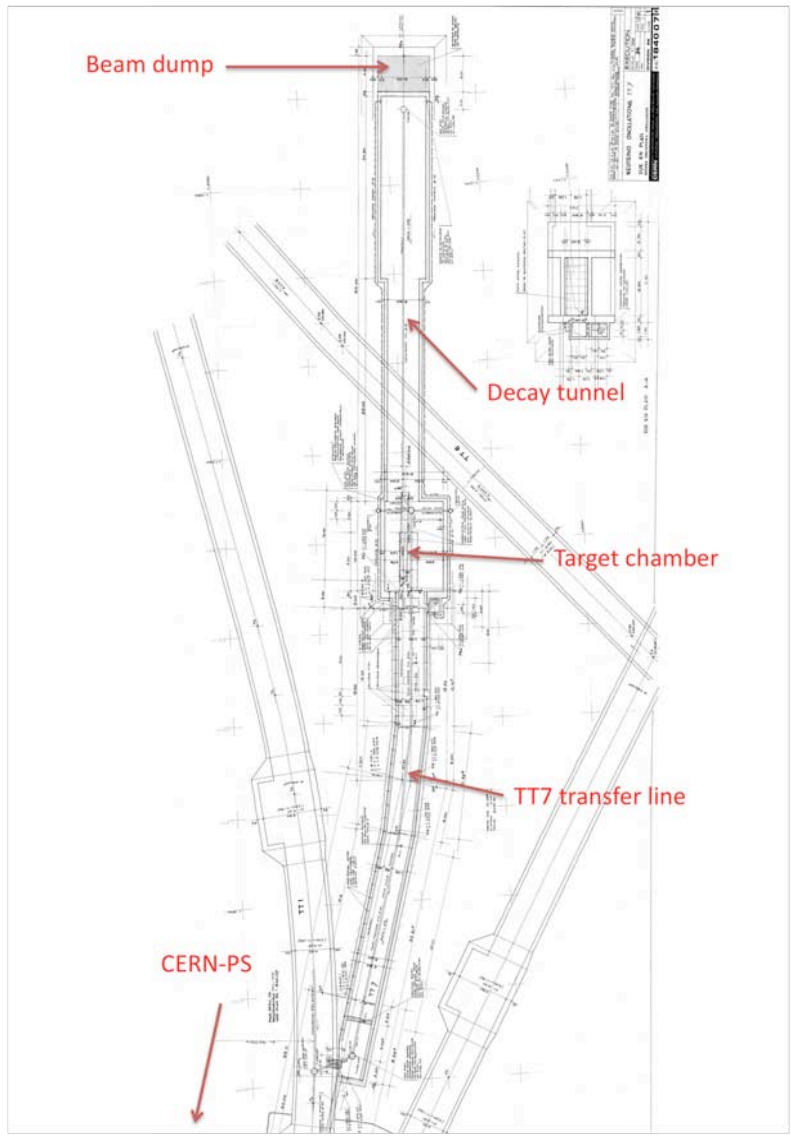

**Figure 5.** Layout of the neutrino beam line from the CERN-PS.

The CERN-PS can deliver routinely on a refurbished TT7 transfer line 3 10$^{13}$ protons per cycle at 20 GeV/c, in the form of 8 equidistant bunches of approximately 60 ns during 2.1 µs. More than 1.25 10$^{20}$ pot/year can be collected by the experiment in one year of data taking for an optimal machine time availability. The actual beam sharing will depend on the extent of requirements from other users, in particular the n-TOF programme, even under the assumption of no major machine upgrade. The proton transfer line, the new target and the horn focussing system must be reconstructed.

As reference case the neutrino beam set-up used by the BEBC-PS180 $\nu_\mu \to \nu_e$ experiment has been chosen. An optimized design of a new and possibly improved beam optics will be the subject of further studies. The 19.2 GeV/c proton beam is extracted from the PS and impinges on a 80 cm long, 6 mm diameter beryllium target. This is followed by a pulsed magnetic horn designed to focus positive particles of momentum around 2 GeV/c into a decay tunnel of about 50 m length. The tunnel cross section is 3.5 x 2.8 m$^2$ for the first 25 m of length and 5.0 x 2.8 m$^2$ for the rest of the length, allowing the decay of mesons with large angular divergence with respect to beam axis in the horizontal plane. The decay tunnel is followed by a 4 m thick iron shield and 65 m of earth to absorb the remaining hadrons and most of the muons (Figure 5).

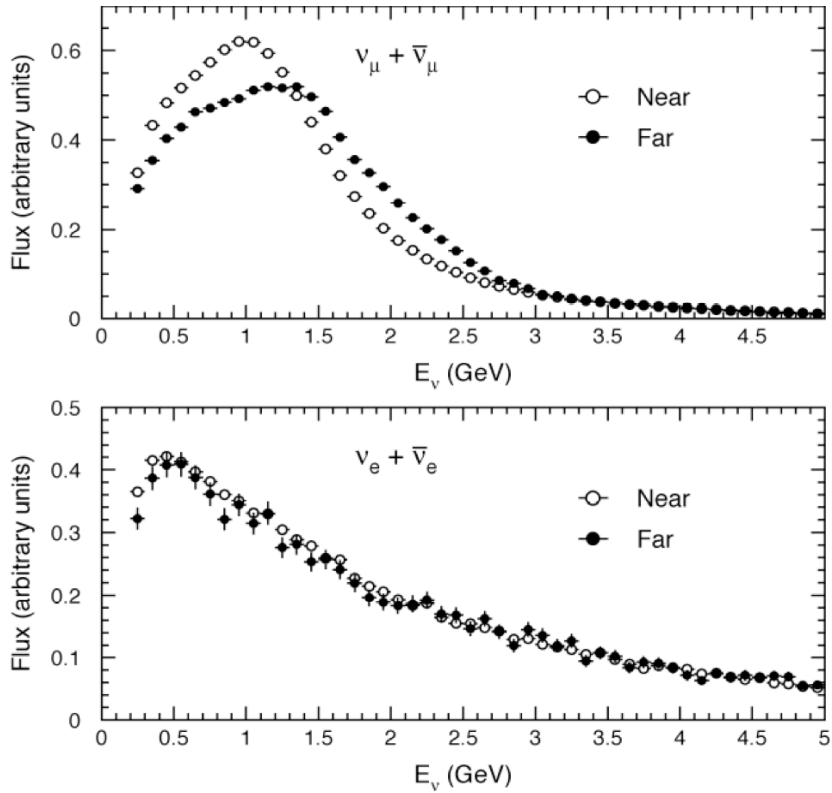

**Figure 6.** CERN-PS neutrino beam energy fluences at the near (127m) and far locations (850m), based on the PS-180 beam line optics, as calculated in [15].

The predicted $\nu_\mu$ and its intrinsic $\nu_e$ contamination spectra for neutrino focusing conditions, as calculated by the I-216/P-311 Collaboration [15], are shown in Figure 6. While the difference between the "near" and "far" positions of the $\nu_\mu$ spectrum are significant, the $\nu_e$ contamination spectra, expected at the level of 0.5% *are very closely identical in the two positions*. The physical reason of this effect has to be identified in the fact that while the $\nu_\mu$ spectrum is dominated by the two body $\pi \to \nu_\mu \mu$ decays, where the neutrino directions are narrowly distributed along the axis, the $\nu_e$ contamination is dominated by the three body decays of K and µ where the neutrino angular spread is much wider (Figure 7). This very relevant property of the beam can be verified measuring experimentally the radial distributions of $\nu_\mu$ and $\nu_e$ neutrino events.

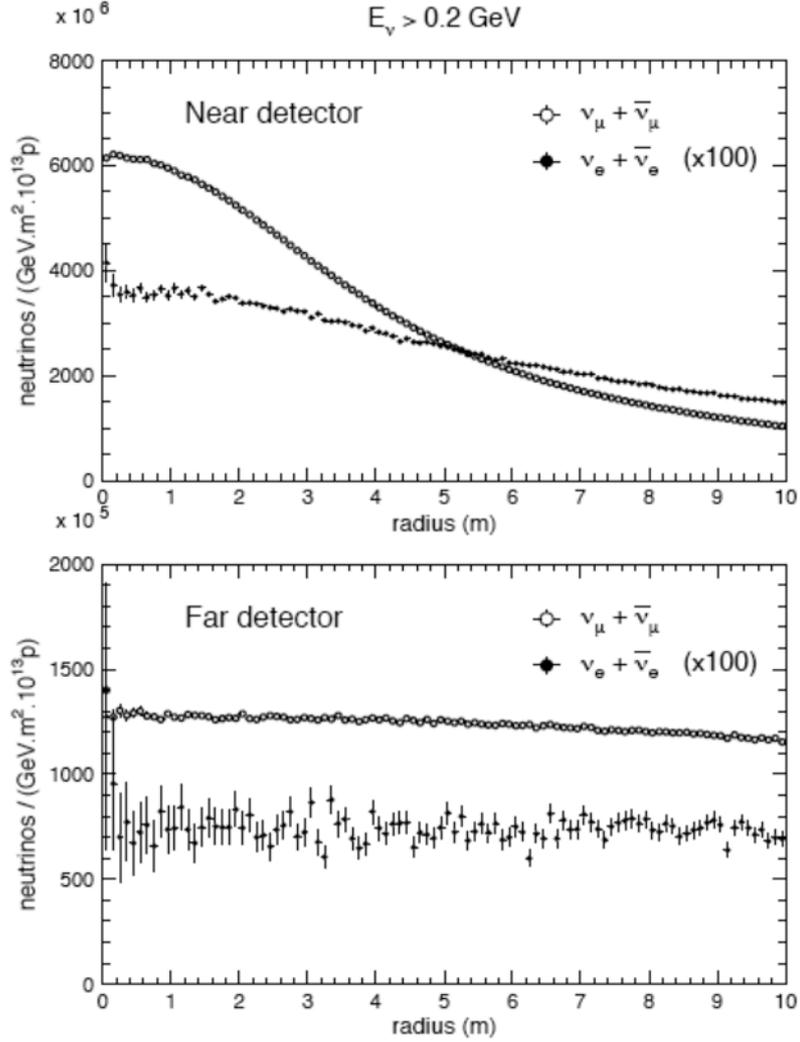

**Figure 7.** CERN-PS neutrino beam radial distributions at the near (127m) and far locations (850m) [15].

Therefore an exact, observed proportionality between the "near" and "far" positions of $\nu_e$ implies directly the absence of neutrino oscillations over the measured interval of L/E.

The sensitivity to $\nu_\mu \to \nu_e$ oscillations can be calculated with the help of the quantity:

$$\Delta_e = \left(\frac{N_e}{N_\mu}\right)^{far} - C_e \left(\frac{N_e}{N_\mu}\right)^{close}$$

where $N_e/N_\mu$ is the ratio of the event number with identified electron and muon as detected in the near and far location and $C_e$ is an energy dependent correction factor that takes into account the difference of the $N_e/N_\mu$ ratio in the two locations. In the energy range below 3.0 GeV, where oscillations are expected, the $C_e$ correction amounts to less than 30 % with the original BEBC-PS180 beam optics [15].

The precision on the determination of $\Delta_e$ is the limiting factor in the sensitivity to neutrino oscillations. In the case of no oscillations the number of electron events $N_e$ is due essentially to the contribution of the genuine electron neutrino interactions. In absence of oscillations $\Delta_e$ is consistent with zero.

Since the un-oscillated neutrino spectra in the close and far detectors have a similar structure, the systematic errors related to the knowledge of the neutrino beam and cross-sections cancel out with this method. A value of $\Delta_e \neq 0$ is therefore the indication for neutrino oscillations. As an example, in Figure 8, the quantity $\Delta_e$ is plotted for two different values of oscillation parameters, normalized to the ratio in the near position ($N_e/N_\mu$).

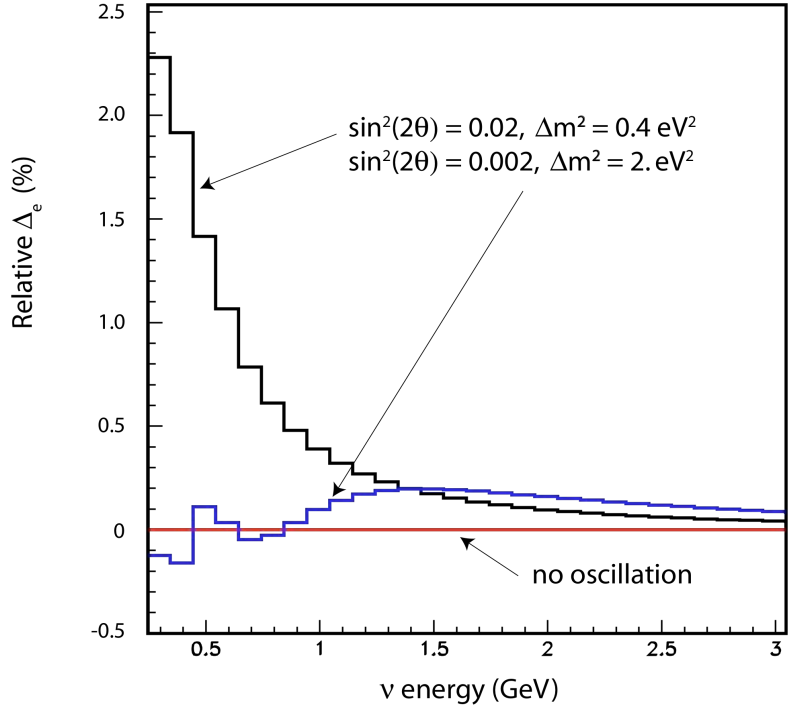

**Figure 8.** Expected behaviour of $\Delta_e$, normalized to the ratio in the near position ($N_e/N_\mu$), for two different values of oscillation parameters.

In Figure 9 a number of experimentally expected oscillation patterns at 850 m are shown for two neutrino (mu - e) oscillations and for some indicative positions of the LSND allowed region (indicated with a star mark). One can see that very different and clearly distinguishable patterns are indeed possible depending on the actual values in the ($\Delta m^2$ – $\sin^2 2\theta$) plane. It appears that the present proposal, unlike LSND and MiniBooNE, can indeed determine in the case of an observed effect, both the mass difference and the value of the mixing angle. In Figure 9 the intrinsic $\nu_e$ background due to the beam contamination is also shown. The magnitude of the expected oscillatory behaviour, although for the moment completely unknown, is in all circumstances well above the background, also considering the very high statistical impact of the experimental measurement.

As an important feature of this LoI, the possibility to run with anti-neutrino beam has been considered in order to further check the LSND claim, which is mainly based on anti-neutrino data. Switching the horn polarity to select negative sign mesons results in an anti-neutrino beam with total fluence of 68 % w.r.t the neutrino mode and a similar energy spectrum.

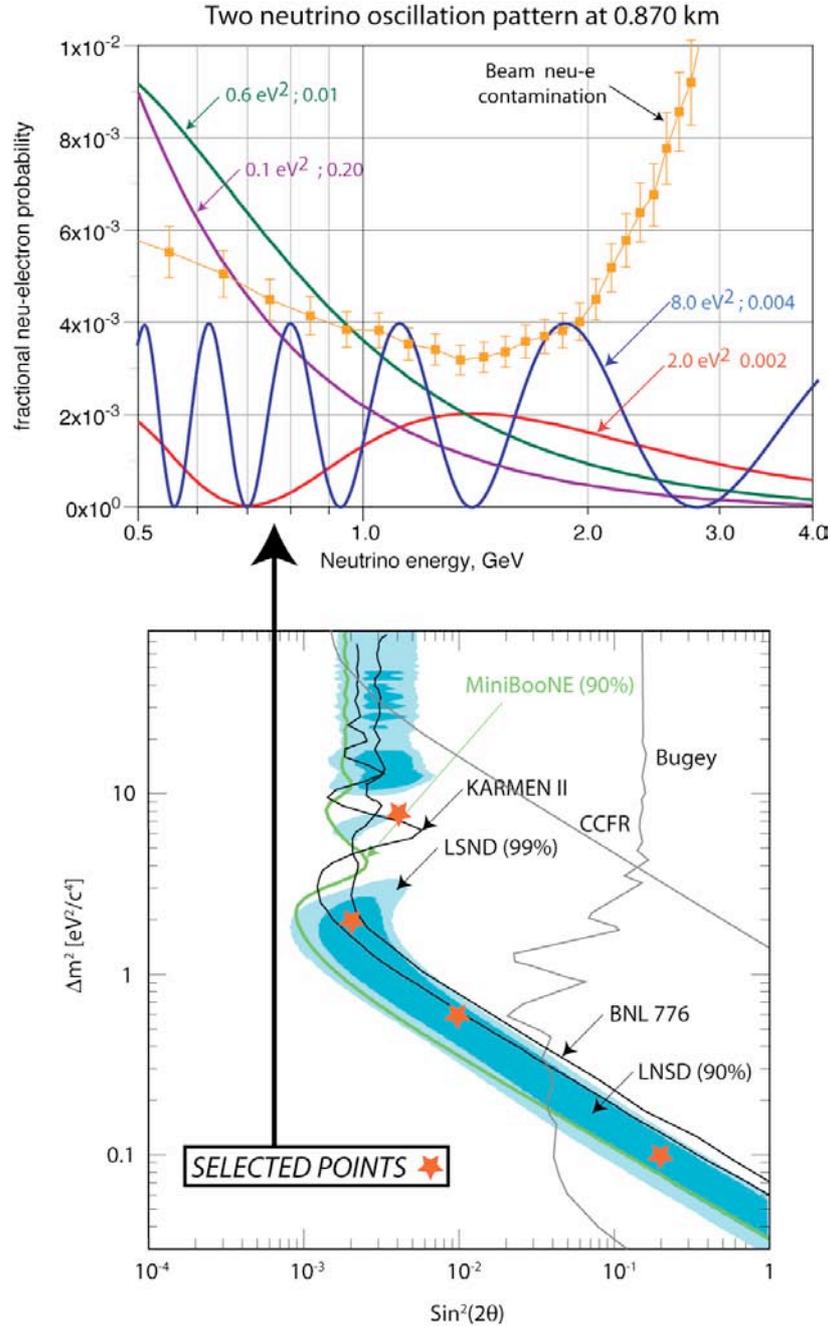

**Figure 9.** The experimentally expected oscillation patterns at 850 m are shown for two neutrino (mu-e) oscillations and for some indicative positions of the LNSD allowed region (indicated with a star mark). The expected background due to the $\nu_e$ beam contamination is also indicated.

**Table 1**. Event rates for the Far and Near detectors given for 2.5 $10^{20}$ pot for $E_\nu < 8$ GeV. Neutrino fluxes are taken from Ref. [15]. The oscillated signals are clustered below 3 GeV of visible energy.

|  | FAR | NEAR |
|---|---|---|
| Fiducial mass | 500 t | 150 t |
| Distance from target | 850 m | 127 m |
| $\nu_\mu$ interactions | 1.2 x $10^6$ | 18 x $10^6$ |
| QE $\nu_\mu$ interactions | 4.5 x $10^5$ | 66 x $10^5$ |
| Events/burst | 0.17 | 2.5 |
| Intrinsic $\nu_e$ from beam | 9000 | 120000 |
| Intrinsic $\nu_e$ from beam ($E_\nu < 3$ GeV) | 3900 | 54000 |
| $\nu_e$ oscillations: $\Delta m^2 = 2.\ eV^2$; $sin^2 2\theta = 0.002$ | 1194 | 1050 |
| $\nu_e$ oscillations: $\Delta m^2 = 0.4\ eV^2$; $sin^2 2\theta = 0.02$ | 2083 | 2340 |

Including anti-neutrino cross-sections, the resulting event rate is 28 % of that collected in neutrino mode. Therefore two years data taking (3.3 $10^5$ $\nu_\mu$ CC in the far detector) will be sufficient to fully explore the LSND allowed region also with $\bar{\nu}_\mu$.

In the LAr-TPC all reaction channels with electron production can be analyzed without the need to restrict the search to the quasi-elastic channel, which accounts for about one half of the events. Events due to neutral currents are also very well identified and can be rejected to a negligible level.

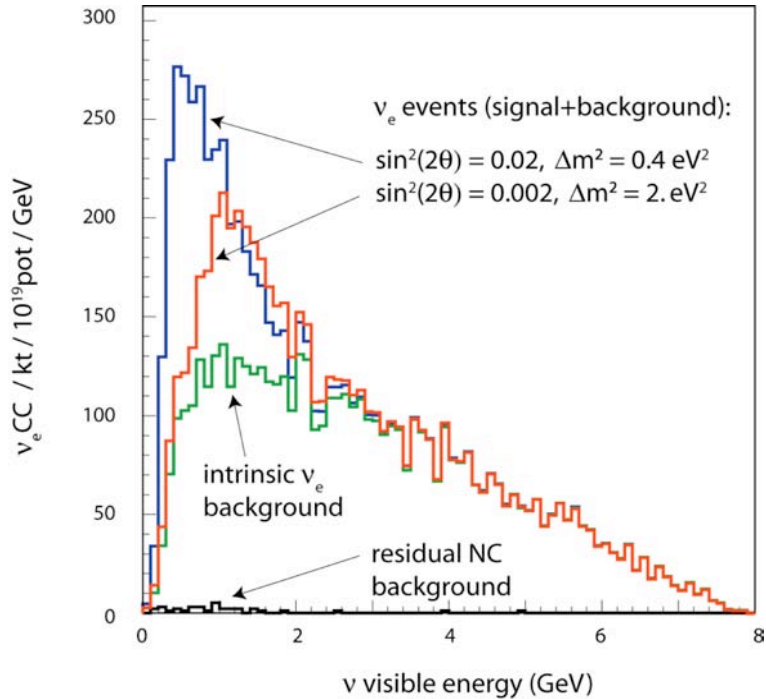

**Figure 10.** Expected excess of $\nu_e$ CC events in the far detector, compared to the $\nu_e$ background (essentially $\nu_e$ from beam). The event sample used to calculate the background and the oscillation signals corresponds to 1.25 $10^{20}$ and 5 $10^{20}$ pot kt statistics respectively.

A full simulation of the expected neutrino events in the LAr detectors has been performed within the FLUKA framework [18], where all interaction processes on Ar nuclei (QE, resonances, DIS) as well as full particle transport are included. It must be stressed that especially at low energies the corrections due to the nuclear effects and Pauli exclusion principle are substantial [19]. However, these effects are the same and therefore cancel out when the spectra in the "near" and "far" positions are compared. The expected numbers of events (with and without $\nu_\mu \leftrightarrow \nu_e$ oscillations) in the CERN-PS beam are shown in Table 1.

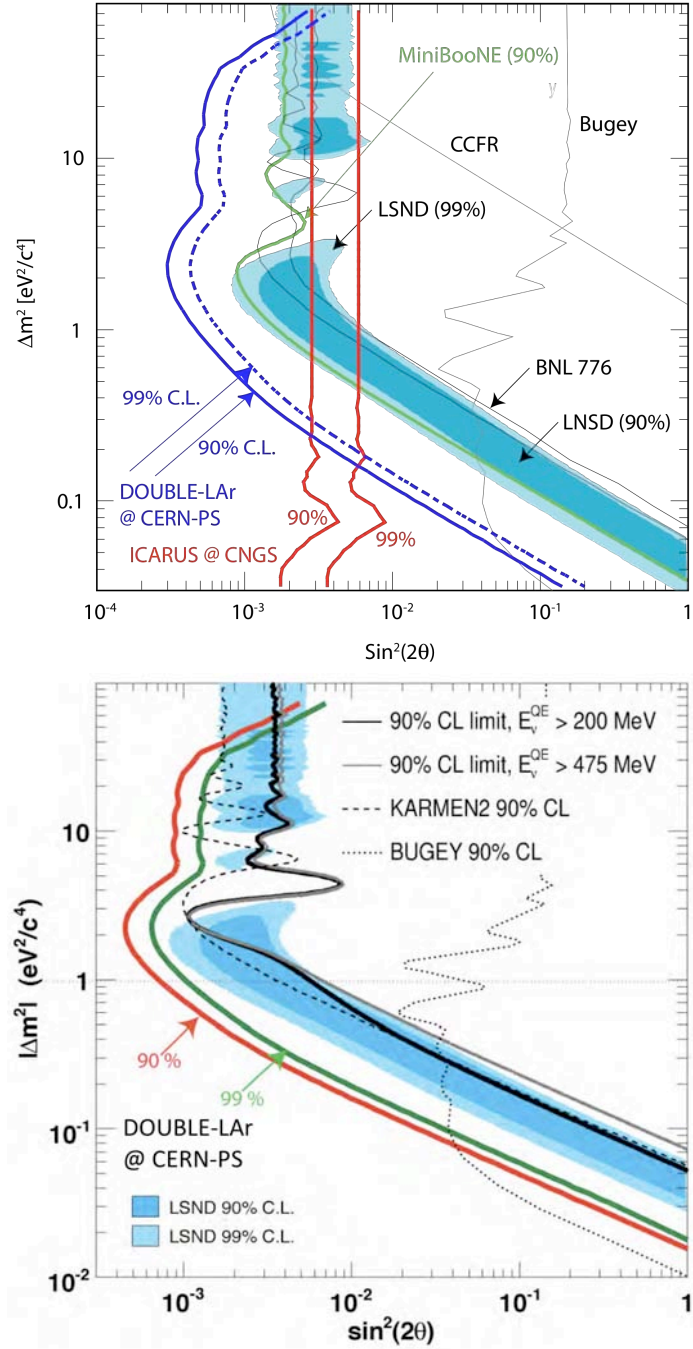

**Figure 11.** Expected sensitivity for the proposed experiment exposed at the CERN-PS neutrino beam (top) and anti-neutrino (bottom) for $2.5 \times 10^{20}$ pot. The LSND allowed region is fully explored both for neutrinos. In the neutrino case, the expectations from CNGS2/ICARUS T600 at LNGS are also shown.

The Monte Carlo simulation shows that the energy reconstruction of electron neutrino interaction events is not affecting the signal/background ratio if a minimal cut of 50 cm in the longitudinal direction and a 10 cm cut on the sides of the sensitive volume are performed. This results in a fiducial volume of about 90 % of the active one. Electron identification is also ensured under these geometrical cuts. Indeed, due to the directionality of the neutrino beam, the probability that an electron escapes from the instrumented volume before initiating a shower is extremely small: only 2 % of the electrons travel through a LAr-TPC thickness smaller than 3 $X_0$ and 0.3 % travel less than 1 $X_0$ in the instrumented volume.

Thanks to the excellent imaging capability of LAr-TPC, $\pi^0$ from $\nu_\mu$ NC events cannot be misidentified as electrons because all events, where both photon conversion points can be distinguished from the $\nu_\mu$ interaction vertex, can be rejected. In the analysis photons converting at more than 2 cm from the $\nu_\mu$ vertex were rejected. The remaining $\pi^0$ background is further reduced by discarding events where the parent $\pi^0$ mass can be reconstructed within 10 % accuracy. Only 3 % of $\pi^0$ survives the cuts for events with vertexes inside the fiducial volume. The remaining photons can be discriminated from electrons on the basis of dE/dX analysis. This method provides a 90 % electron identification efficiency with photon misidentification probability of 3 %. The final $\pi^0$ mis-interpretation probability is 0.1 %, the corresponding electron neutrino detection efficiency is 90 % within the fiducial volume (Figure 10).

In Figure 11, the expectations for two year of data taking is shown both for the neutrino and anti-neutrino cases.

Precise measurements of the neutrino cross-section in the 0-5 GeV energy range are required by present and future neutrino oscillation experiments. Existing data on charged current quasi-elastic, deep inelastic, and single pion production are affected by a large uncertainty, especially at the lower energies [20]. Owing to the very low detection threshold of the Liquid Argon technique, the exposure at the CERN-PS neutrino beams could significantly improve the neutrino cross-section knowledge. Approximately $1.2\ 10^6$ and $1.8\ 10^7$ charged current events per $2.5\ 10^{20}$ pot will be recorded from the CERN-PS in the far and near detectors respectively, with a neutrino spectrum peaked at around 1 GeV. Neutral current cross-sections are also measured, with approximately $4.4\ 10^5$ and $6.0\ 10^6$ events in the far and near detectors respectively.

## 3. The Liquid Argon detectors

The ICARUS T600 detector [13] is the largest liquid Argon TPC ever built, with a size of about 600 t of imaging mass. The design and assembly of the detector relied on industrial support and represents the application of concepts matured in laboratory tests to the kton-scale .

The operational principle of the LAr TPC is based on the fact that in highly purified LAr ionization tracks can be transported practically undistorted by a uniform electric field over macroscopic distances. Imaging is provided by a suitable set of electrodes (wires) placed at the end of the drift path, continuously sensing and recording the signals induced by the drifting electrons. Non-destructive read-out of ionization electrons by charge induction allows detecting the signal of electrons crossing subsequent wire planes with the wires running along different orientation. This provides several projective views of the same event, hence allowing space point reconstruction and precise calorimetric measurement.

The ICARUS T600 LAr detector consists of a large cryostat split into two identical, adjacent half-modules, with internal dimensions $3.6 \times 3.9 \times 19.6$ m$^3$ each (see Figure 12). Each half-module houses two Time Projection Chambers (TPC) separated by a common cathode, a field shaping system, monitors and probes, and two arrays of photo-multipliers, coated with TPB waveshifter. Externally the cryostat is surrounded by a set of thermal insulation layers. The detector layout is completed by a cryogenic plant made of a liquid Nitrogen cooling circuit, maintaining the LAr temperature uniform, and by a system of LAr purifiers. Each TPC is made of three parallel planes of wires, 3 mm apart. The first faces the drift region, with horizontal wires, the wires of the other two planes are at ± 60° from the horizontal direction. By appropriate voltage biasing, the first two planes (Induction planes) provide signals in non-destructive way, whereas the charge is finally collected in the last one (Collection plane). The maximum drift path, i.e. the distance between the cathode and the wire planes, is 1.5 m and the nominal drift field 500 V/cm. The total number of wires in the T600 detector is about 53000. The signals coming from each wire are independently digitized every 400 ns. The electronics was designed to allow continuous read-out, digitization and independent waveform recording of signals from each wire of the TPC. The measurement of the time of the ionizing event, the "$T_0$ time" which can be determined via the prompt scintillation light produced by ionizing particles in LAr, together with the knowledge of the electron drift velocity, provides the absolute position of the tracks along the drift coordinate.

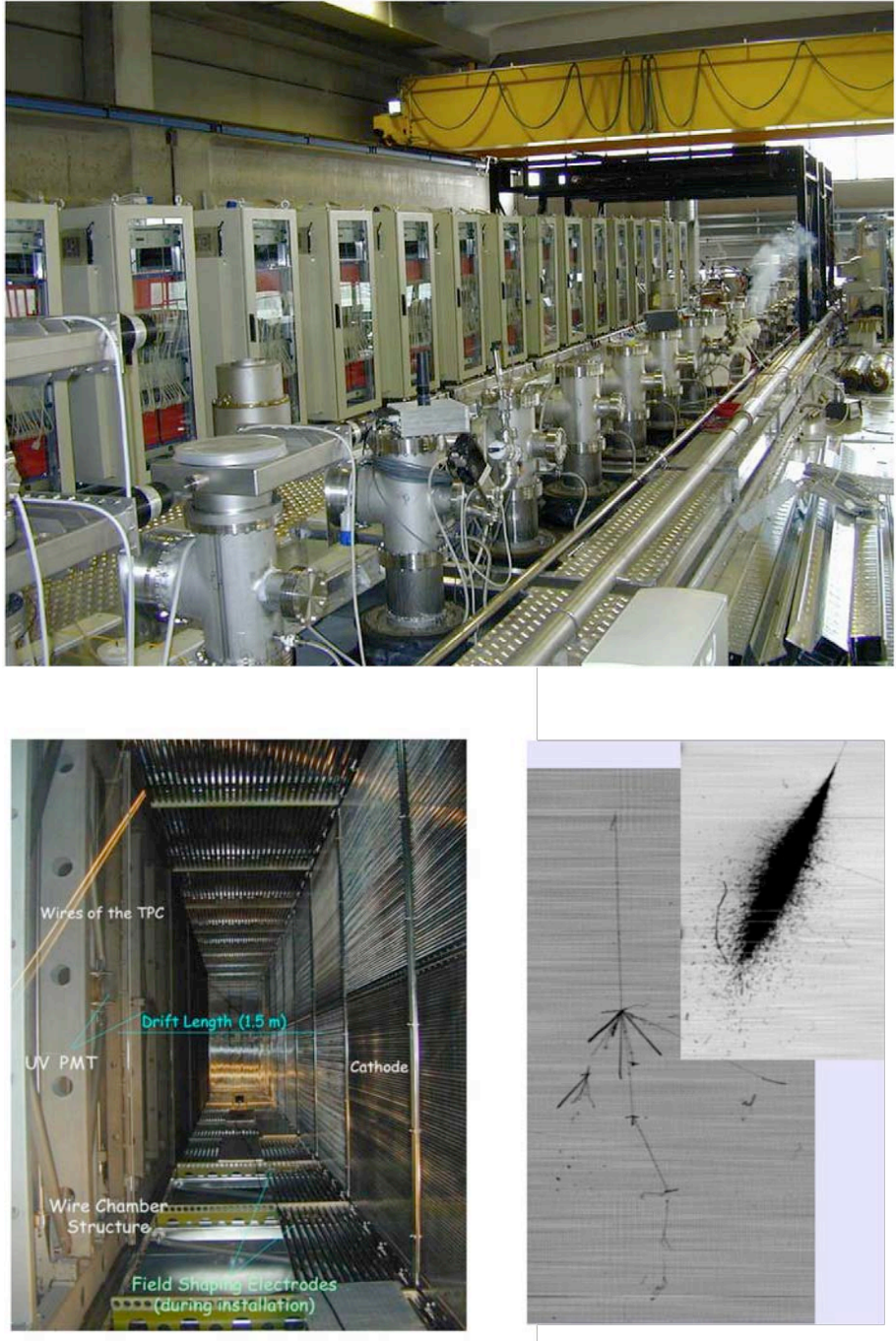

**Figure 12.** Top: The top of one T300 module in Pavia showing the disposition of the electronic racks for the left chamber and of the chimneys with the feedthrough flanges for the right chamber. Bottom-Left: view of the one of the drift region of the T600 LAr-TPC. Bottom-Right: cosmic ray events recorded with the T600 in Pavia.

The ICARUS detector provides a measurement of ionization energy loss of a track similarly to a bubble chamber. A complete three dimensional spatial and calorimetric picture of events is built by extracting the physical information contained in the wire output signals, i.e.

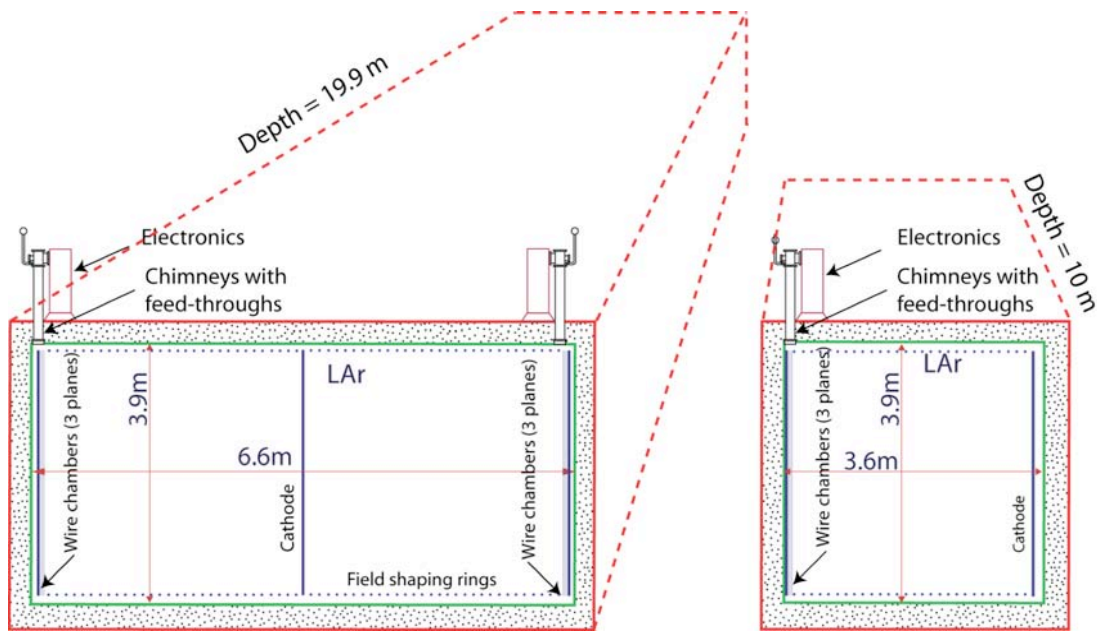

**Figure 13.** Artistic view of the proposed layout of the far (left) and near (right) LAr-TPC.

the energy deposition by the different particles and their hit position in the LAr. The measurement of the dE/dx and the related position for a large number of points along the track allow estimating the particle momentum, from range (for stopping particles) or multiple scattering, and the particle identification.

The proposed far detector closely resembles the T600 module maintaining the size of the external dewar and the mechanical design for a total LAr mass of about 600 t (see Figure 13-left). The LAr container is of parallelepiped shape with internal dimensions of 6.6 x 3.9 x 19.9 m$^3$. The basic elements of the cold vessels are panels with a thickness of 150 mm and a surface of 2.0 x 3.9 m$^2$ in the case of side walls and a surface of 2.0 x 6.9 m$^2$ in the case of top and bottom walls. They are made of aluminum honeycomb, surrounded by frames of aluminum profiles and sandwiched between two aluminum skins. The two end caps (6.6 x 3.9 m$^2$) both formed by two special panels joined by an H-shaped profile, are added at the end. Holes for the way out of signal cables of the internal detectors and control instrumentation are located on the ceiling of the container, in correspondence with the two frames of the wire chambers. The ports for LAr filling, gaseous argon (GAr) recirculation and other services like vacuum instrumentation and safety disks, as well as two flanges (manholes) of 500 mm diameter used for inspection and for the final assembly operations, are also located on the top side.

The thermal insulation of the detector is ensured by an additional layer of passive insulation material, corresponding to a heat loss of less than 10 W/m$^2$, or about 6 kWatt of cold power.

Cooling of the cryostat will be performed in analogy to T600 by circulating pressurized LN$_2$ at 2.7 bar abs and nominal temperature of 89 K inside the cooling circuits of the two LAr containers. The circulation speed is defined by the request to maintain a maximum thermal gradient lower than 1 K on the nitrogen circuit. To ensure an acceptable initial LAr purity, the detector can be evacuated before filling.

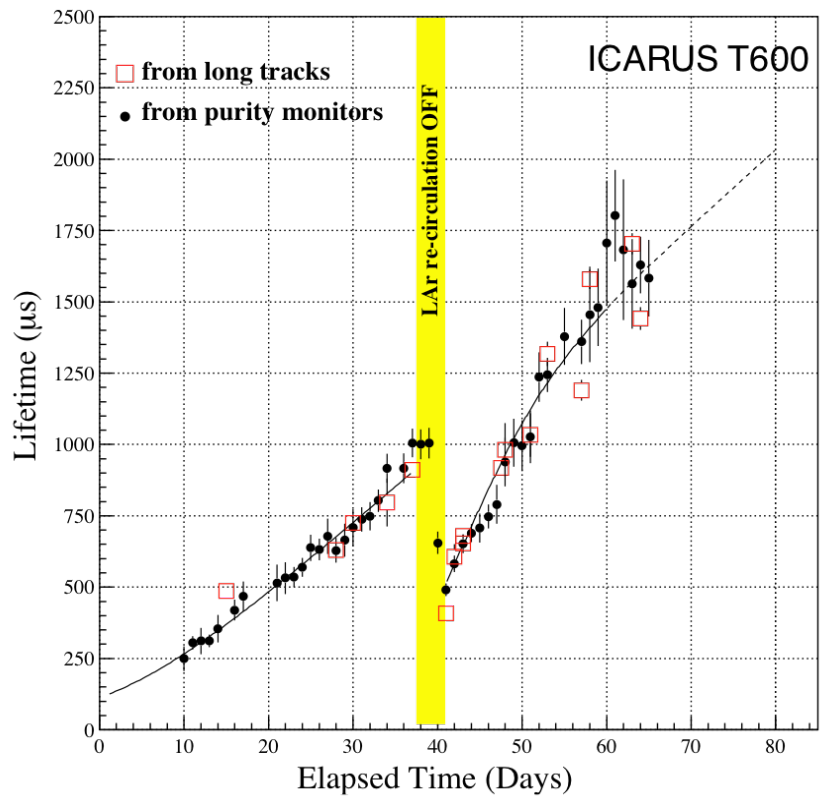

**Figure 14.** Evolution of the free electron lifetime during the T600 test run in Pavia from purity monitor (dots) and from ionizing tracks analysis (squares). A period when the liquid re-circulation was switched off is indicated. The solid line shows a fit result [16].

It will be equipped with gas and liquid recirculation systems, both required to attain a very high free-electron lifetime (few ms) in less than one month. The continuously active gas recirculation units collect the gas from the chimneys that host the cables for the wire chamber read-out. The gas is re-condensed into a LN$_2$ re-condenser with the liquid dropping into an Oxysorb/Hydrosorb filter placed below the re-condenser. The purified argon is sent back into the LAr container just below the LAr surface.

The main purpose of the liquid recirculation (circulating ~ 2 m$^3$/h) is to purify the argon just after filling and until the required LAr purity is reached. In addition, it has to restore the purity in case of accidental pollution during the detector run, e.g. due to a sudden depressurization of the gas volume. Liquid recirculation units consist of an immersed,

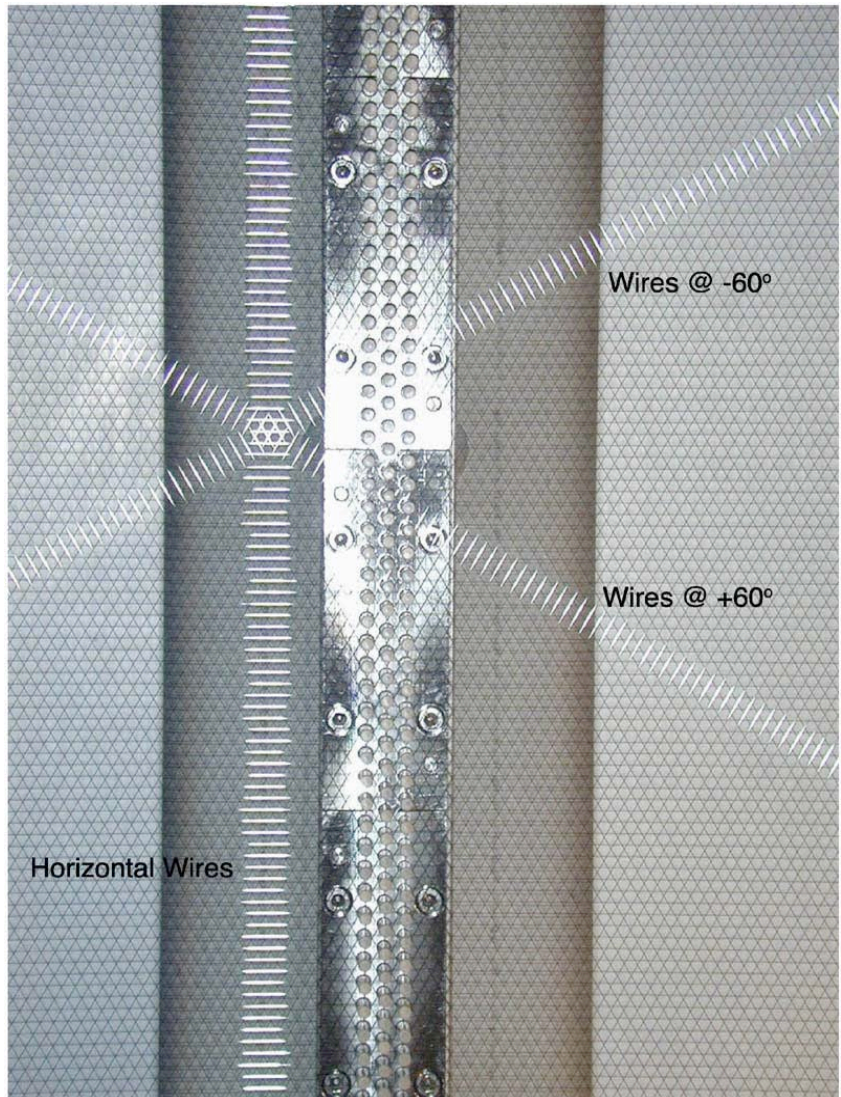

**Figure 15.** The three wire planes of the LAr-TPC installed in the T600 detector.

cryogenic, liquid transfer pump placed inside an independent dewar.

Internally, the drift distance could be extended to more that 3 m, twice that of ICARUS T600, in order to simplify the structure of the LAr-TPC. The experience with the T600 shows that the required LAr purity level is achievable with the standard ICARUS purification system (Figure 14) [16]. At the same time it has been demonstrated that doubling the high voltage on the cathode is entirely feasible [21]. The cathode plane is central with two sets of 3 anodic wire planes at the opposite sides of the cryostat. Both the cathode and wire chambers will be identical to that of the T600 detector. As a consequence, the number of the wire chambers is reduced by a factor two as well as the number of the read-out electronic channels.

Wires are made of AISI 304V stainless-steel with a wire diameter of 150 µm: the length of the wires ranges from 9.42 to 0.49 m depending on the TPC plane type and on the actual position of the wire in the plane itself (Figure 15). The wire-frame design is based on an innovative concept already used in T600: the variable geometry design (weight bridge). The beams of the wire frame are, in fact, movable. The upper and lower horizontal beams are rigidly connected to each other on the back of the frame by a set of calibrated springs, while the vertical beams are connected to the sustaining structure also by springs (Figure 16). This allows: (1) to set the tension of the wires after easy installation, (2) to compensate for possible overstress during the cooling and LAr filling phases and (3) to counteract the flexibility of the structure. This design was successfully adopted in ICARUS T600 for more that 53000 wires, none of which broke during testing in Pavia and moving the detector to LNGS.

The TPC signals are extracted with the same technique as in T600, using the well-experienced feed-through system. Also the electronic chain is inherited from the ICARUS T600 detector as well as the DAQ system, with a possible evolution based on an upgraded DAQ scheme that implements the same ICARUS T600 architecture with more performing new components and different modularity [21].

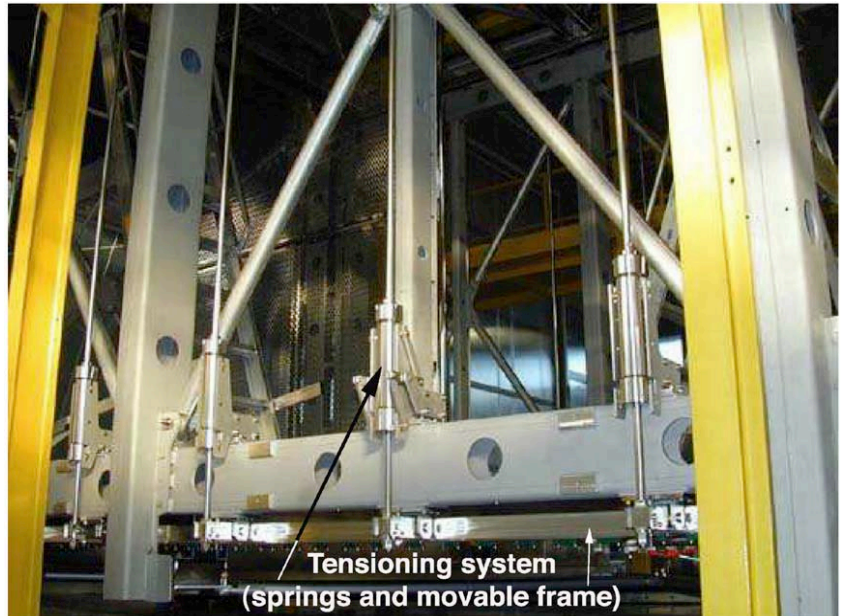

**Figure 16.** Detail of the wire tensioning mechanics: a 2 m long portion of the wire frame equipped with three tensioning springs.

The DAQ system is based on a low noise analogue front-end, with J-fet input stage, followed by a multiplexed 10-bit ADC and by a digital VME module that performs local storage and data compression. It has been tested satisfactorily in the 2001 run. The upgraded scheme will be based on a hybrid sub-module integrating several channels of amplification and possibly the analogue to digital converter. A revision of the ICARUS analogue electronics is underway with the aim to further improving the front-end performance [21].

A novel technique for the realization of feed-through's has been developed. INFN holds a patent (RM2006A000406) for this technology that allows easy realization of feed-through's with high-density vias and different shapes.

As in the T600, a set of few tens of PMT's, coated with TPB waveshifter, will be placed in LAr behind the wire chambers (Figure 17) to detect the prompt scintillation light for triggering purposes. The self-triggering system will be based on both the PMT and wire signals as it is foreseen in the T600 readout.

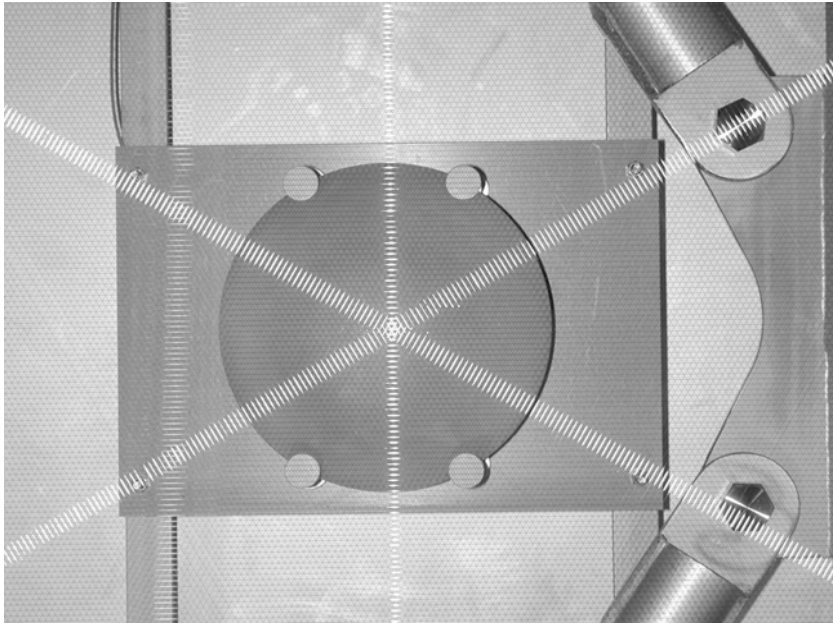

**Figure 17.** One of the PMTs positioned behind the three wire planes of the LAr-TPC

The near detector design (see Figure 13-right) will be similar to the one of the T300 semi-module but with the length reduced by a factor 2, i.e. 10 m. for a total LAr mass of 150 t, the cathode displaced to one side and just one set of wires at the opposite side. This solution permits to use the same mechanics of the wire chambers of the far detector.

This design strategy allows exploiting the already developed and tested technology of ICARUS T600 without the need of any further R&D studies. Finally, the reconstruction and analysis tools developed for ICARUS to study neutrino physics both with cosmic rays and the CNGS beam, could also be exploited by the proposed experiment.

## 4. Conclusions

The exposure to a low energy neutrino beam at CERN-PS of two LAr-TPC's, similar in size and design to the ICARUS T600, will allow investigating the existence of sterile neutrinos through the measure of $\nu_\mu \to \nu_e$ oscillations.

The use of two LAr-TPC's at a near and a far location enables to minimize the effects of systematic uncertainty related to the neutrino beam and to the neutrino cross-sections. The quality of the LAr-TPC as tracking and calorimetric detector is essential to reject at negligible level the contribution of neutral currents in the $\nu_e$ background. A sensitivity of $\sin^2 2\theta < 3\ 10^{-4}$

– 21 –

(for $\Delta m^2 < 2$ eV$^2$) and $\Delta m^2 < 0.02$ eV$^2$ (for $\sin^2 2\theta = 1$) at 90 % C.L. is expected with a two year exposure at the CERN-PS $\nu_\mu$ beam. A similar sensitivity can be reached with $\bar{\nu}_\mu$ beam.

The relevant neutrino and antineutrino cross-sections in the energy range 0-3 GeV will be carefully measured in LAr. Exploring this energy domain is a necessary prerequisite for the future long baseline experiments aiming at the determination of the $\theta_{13}$ mixing angle and a search for CP-violation in the leptonic sector.

The experience of the ICARUS T600 on surface makes us confident that the presently proposed detector could be constructed in a relatively short time and will feature the required characteristics and performance.

**References**

[1]   A. Aguilar et. al. [LSND coll.], Phys. Rev. D 64 (2001) 112007.

[2]   C. Athanassopoulos et al. [LSND coll.], Phys. Rev. Lett. 81 (1998) 1774-1777.

[3]   M. Sorel, J.M. Conrad, M.H. Shaevitz, Phys. Rev. D70 (2004) 073004;

P.Q. Hung, arXiv:hep-ph/0010126;

D.B. Kaplan, A.E. Nelson, N. Weiner, Phys. Rev. Lett. 93 (2004) 091801;

V. Barger, D. Marfatia, K. Whisnant, Phys. Lett. B576 (2003) 303;

G. Barenboim, N.E. Mavromatos, Phys. Rev. D70 (2004) 093015;

V.A. Kostelecky, M. Mewes, Phys. Rev. D70 (2004) 076002;

T. Katori, V.A. Kostelecky, R. Tayloe, , arXiv:hep-ph/0606154;

H. Pas, S. Pakvasa, T.J. Weiler, Phys. Rev. D72 (2005) 095017;

S. Palomares-Ruiz, S. Pascoli, T. Schwetz, JHEP 509 (2005) 48;

S. Palomares-Ruiz, JoP Conf. Series 39 (2006) 307.

[4]   A.A. Aguilar-Arevalo et al., Phys. Rev. Lett. 98 (2007) 231801.

[5]   A. A. Aguilar-Arevalo et al., Phys.Rev.Lett. (in print)

[6]   A. A. Aguilar-Arevalo et al., Phys. Rev. Lett. 102 (2009) 101802.

[7]   C. Rubbia for the ICARUS Coll., CNGS2 (ICARUS), presented at the 78th meeting of the SPSC, CERN, 3 october 2006

[8]   H. Chen et al. [MicroBooNE coll.], A Proposal for a New Experiment Using the Booster and NuMI Neutrino Beamlines: MicroBooNE, FNAL Proposal (2007) and Addendum (2008).

[9]   I. Stancu et al. [OscSNS Coll.], The OscSNS White Paper (2008) [http://physics.calumet.purdue.edu/~oscsns].

[10]  G. Acquistapace et al., The CERN Neutrino Beam to Gran Sasso, CERN 98-02, INFN/AE-98/05 (1998); CERN-SL/99-034(DI), INFN/AE-99/05 Addendum;

Ferrari et al., Nucl. Phys. Proc. B 145(Suppl.) (2005) 93.

[11]  M. Maltoni et al., Phys. Rev. D76 (2007) 093005.

[12] C. Rubbia, The Liquid-Argon Time Projection Chamber: A New Concept For Neutrino Detector, CERN-EP/77-08 (1977); ICARUS Coll., ICARUS initial physics program, ICARUS-TM/2001-03 LNGS P28/01 LNGS-EXP 13/89 add.1/01; ICARUS Coll., Cloning of T600 modules to reach the design sensitive mass, ICARUS-TM/2001-08 LNGS-EXP 13/89 add.2/01.

[13] S. Amerio et al. [ICARUS Coll.], Nucl. Instr. And Meth A527 (2004) 329.

[14] C. Angelini et al. [PS180 Coll.], Phys. Lett. B 179 (1986) 307.

[15] M. Guler et al. [I216/P311 Coll.], Search for $\nu_\mu \leftrightarrow \nu_e$ oscillation at the CERN PS (Proposal), CERN-SPSC/99-26, SPSC/P311, 1999.

[16] S. Amoruso et al. [ICARUS Coll.] Nucl. Instrum. Meth. A 523 (2004), 275;

S. Amoruso et al. [ICARUS Coll.] Nucl. Instrum. Meth. A 516 (2004) 68;

S. Amoruso et al. [ICARUS Coll.] Eur. Phys. J. C33 (2004) 233;

Ankowski et al. [ICARUS Coll.] Eur. Phys. J. C48 (2006) 667.

[17] F. Arneodo et al. [ICARUS-Milano Coll.], Phys. Rev. D 74 (2006) 112001.

[18] G. Battistoni et al., "The FLUKA code: Description and benchmarking", AIP Conf. Proc. 896, 31-49, (2007);

G. Battistoni et al., A neutrino-nucleon interaction generator for the FLUKA Monte Carlo code, Proceedings of the 12th International conference on nuclear reaction mechanisms, Varenna (Italy), June 15 - 19, 2009, in press;

Ferrari et al., FLUKA, a multi particle transport code (program version 2005), CERN-2005-10, INFN/TC-05/11, 2005.

[19] G. Battistoni et al., Acta Phys. Polon. B37 (2006) 2361.

[20] G.P. Zeller, Neutrino Cross Sections: Past, Present and Future, talk at NuInt07, FNAL (2007) [http://theory.fnal.gov/jetp/talks/zeller.pdf];

G.P. Zeller, Low Energy Neutrino Cross Sections: Comparison of various Monte Carlo Predictions to Experimental Data, talk at NuINt02 Irvine (2002) [hep-ph/0312061v1].

[21] D. Angeli et al., JINST 4 P02003 (2009).